\documentclass[aps,prl,twocolumn,superscriptaddress,amsmath,amssymb,showpacs]{revtex4-1}
\usepackage{graphicx}
\usepackage{graphics}
\usepackage{amsfonts,epsfig}
\usepackage{float}
\usepackage{color}
\allowdisplaybreaks

\begin{document}
\title{Fast magnetic field manipulations and nonadiabatic geometric phases of nitrogen-vacancy center spin in diamond}

\author{Wen-Qi Fang}
\affiliation{Beijing National Laboratory for Condensed Matter Physics, Institute of Physics, Chinese Academy of Sciences, Beijing 100190, China.}
\affiliation{School of Physical Sciences, University of Chinese Academy of Sciences, Beijing 100190, China}
\author{Bang-Gui Liu}%
 \email{bgliu@iphy.ac.cn}
\affiliation{Beijing National Laboratory for Condensed Matter Physics, Institute of Physics, Chinese Academy of Sciences, Beijing 100190, China.}
\affiliation{School of Physical Sciences, University of Chinese Academy of Sciences, Beijing 100190, China}

\date{\today}

\begin{abstract}
Fast quantum spin manipulation is needed to design spin-based quantum logic gates and other quantum applications. Here, we construct exact evolution operator of the nitrogen-vacancy-center (NV) spin in diamond under external magnetic fields and investigate the nonadiabatic geometric phases, both cyclic and non-cyclic, in these fast-manipulated NV spin systems. It is believed that the nonadiabatic geometric phases can be measured in future experiments and these fast quantum manipulations can be useful in designing spin-based quantum applications.
\end{abstract}

\maketitle


{\it Introduction.} Analytically solvable quantum  dynamical systems have been sought since the birth of quantum mechanics. Perhaps, the most famous examples include Landau-Zener model\cite{Landau, Zener}, Rabi problem\cite{Rabi}, and hyperbolic secant pulses\cite{Rosen}. These still make an active area in recent years\cite{zhou, Braak, Econo}, and play an important role in quantum control and quantum computation field\cite{Petta, Golter, Greilich} because analytical solutions are often useful in designing fast, precise, and noise resistant pulses\cite{Econo, Economou, Motzoi}. In a recent work\cite{Barnes}, a systematic method was proposed to construct numerous analytical solutions, and more time-dependent two-level solvable Hamiltonians were presented\cite{Barnes1, Messina}. This analytical method is also very powerful in designing control pulses to realize robust and fast quantum gates\cite{Xin, Sophia, Fang}.

In addition, using such methods, we can study nonadiabatic (and even non-cyclic) geometric phase\cite{Pancharatnam,Berry,AA,Samuel,Mukunda} of these dynamical systems.
The geometric phase is a consequence of quantum kinematics and is thus independent of the detailed nature of the dynamical origin of the path in state space. Thanks to this particular property, all-geometric approach, holonomic quantum computation\cite{Zanardi, Pachos}, can be used  as a tool to achieve fault-tolerance\cite{Duan, Wang} and robust quantum processing\cite{Vedral, Sjo}. Geometric quantum computation were proposed in many systems, such as trapped ions or atoms\cite{Duan, Recati}, superconducting qubits\cite{Faoro}, and quantum dots\cite{Solinas}, and afterward were already realized in several experiments\cite{Abdumalikov, Feng, Toyoda}. Recently, high-fidelity realization of quantum gates are realized by an individual solid-state spin\cite{CZu, Silvia} based on proposal of non-Abelian holonomic quantum computation\cite{Sjoq, Johansson}. More detailed information about geometric phases in quantum information can be found in a recent review article\cite{S}. Here, we will investigate the nonadiabatic geometric phase of the NV-center spin system in diamond during its fast time evolution.


{\it Exact evolution of quantum states.} The Hamiltonian of the NV center spin $\vec{S}$, in the presence of time
dependent magnetic field $\vec{B}(t)$=$(B_x(t),B_y(t),B_z(t))$, can be written as
\begin{equation}\label{eq1}
H=DS_{z}^{2}+\gamma \vec{S}\cdot\vec{B}(t),
\end{equation}
where $\hbar=1$ is used, $D=2.87$GHz is the zero-field splitting,
and $\gamma=2.8$MHz/G is the electron gyromagnetic ratio. Accordingly, the
Schr\"odinger equation for time-evolution operator $U$ is given by
$i\frac{d}{dt}U=HU$.

By choosing $|\pm \rangle$, defined as $(|+1\rangle \pm|-1\rangle)/\sqrt{2}$, as the qubit basis of the NV center spin, we can obtain exact evolution operator of the NV center spin under time-dependent magnetic field by mapping the three-level system of the NV center spin on a two-level system under a time-dependent magnetic field\cite{Fang} and using the existing exact analytical results of the quantum two-level system\cite{Barnes,Barnes1}. However, we notice that with Barnes's method\cite{Barnes1}, it is difficult to obtain physically reasonable pulse which is zero in amplitude at initial and ending times. So we use Messina's method\cite{Messina} to construct evolution operator. Actually, it is much easier to construct the essential part in this method, although the two methods are essentially equivalent to each other. With special magnetic field $\vec{B}(t)$=$B_0(t)(\alpha,\beta,0)$, the Hamiltonian Eq. (\ref{eq1}) in the new basis of ($|+\rangle$,$|0\rangle$,$|-\rangle$) can be expressed as
\begin{equation}\label{eq2}
H_{o}=\left(\begin{array}{ccc}
D & \alpha\gamma B_0 & 0 \\
\alpha\gamma B_0 & 0 & i\beta\gamma B_0 \\
0 & -i\beta\gamma B_0 & D \\
\end{array}\right).
\end{equation}
Then, we can easily derive its exact evolution operator\cite{Fang}:
\begin{equation}\label{eq3}
\begin{split}
&U_{o}(\theta,t)=d(t)\times \\
&\left(
\begin{array}{ccc}
\bar{u}_{11}\alpha^2+d(t)\beta^2 & -\bar{u}_{21}^{*}\alpha &
-i\left(d(t)-\bar{u}_{11}\right)\alpha\beta \\
\bar{u}_{21}\alpha & \bar{u}_{11}^{*} & i\bar{u}_{21}\beta \\
i\left(d(t)-\bar{u}_{11}\right)\alpha\beta & i \bar{u}_{21}^{*}\beta
& d(t)\alpha^2+\bar{u}_{11}\beta^2 \\
\end{array}
\right),
\end{split}
\end{equation}
where $d(t)$ is defined as $e^{-i\frac{D}{2}t}$, and $\alpha$ and $\beta$ can be parameterized as $\alpha=\cos \theta$ and $\beta=\sin \theta$ ($-\pi \le \theta \le \pi$), respectively. The matrix elements
$\bar{u}_{11}$ and $\bar{u}_{21}$ are given by
\begin{equation}
\left\{\begin{array}{c}
\bar{u}_{11}(t)=\Re({u_{11}})+i \Im({u_{21}})\\
\bar{u}_{21}(t)=i\Im({u_{11}})+\Re({u_{21}})\\
\end{array}\right.
\end{equation}
where $\Re$ and $\Im$ represent the real and imaginary parts, respectively.
The explicit expressions for the parameters $u_{11}$ and $u_{21}$  and the time
dependent magnetic field are given by
\begin{equation}\label{eq4}
\left\{\begin{array}{l}
{u_{11}}(t)=\cos (\frac{D}{2}W_c(\Theta,t))\exp\{-\frac{i}{2}[\Theta(t)-W_s(\Theta, D, t)]\} \\
{u_{21}}(t) = -i\sin(\frac{D}{2}W_c(\Theta,t))\exp\{\frac{i}{2}[W_s(\Theta, D, t)-\Theta(t)]\} \\
\gamma B_0(t) = \frac{\dot{\Theta}}{2}+\frac{D}{2}\sin \Theta \cot (DW_c(\Theta,t))
\end{array}\right.
\end{equation}
It should be noted that here we require $\Theta(0)=0$ and use the definitions:
\begin{equation}\label{eqadd1}
\left\{\begin{array}{l}
W_c(\Theta,t)=\int_{0}^{t}\cos \Theta(t^{'})dt^{'}\\
W_s(\Theta, x, t)=\int_{0}^{t}\frac{x\sin \Theta(t^{'}) }{\sin(xW_c(\Theta,t^{'}))}dt^{'}.
\end{array}\right.
\end{equation}
From above equations (\ref{eq3}) and (\ref{eq4}), we can see that
evolution operator(\ref{eq3}) is dependent on the magnetic field direction characterized
by parameter $\theta$. If the time-dependent function $\Theta(t)$ is given,
the evolution operator (\ref{eq3}) will be determined immediately. Using this time evolution operators, we can manipulate the single NV center spin, exactly and efficiently.

Experimentally, the NV center spin can be easily prepared in state
$|0\rangle$. We try to realize state transfer between $|0\rangle$ and $|\pm\rangle$.
With the time evolution operator $U_{o}$ applied, the state $|0\rangle$ will become
\begin{equation}\label{eq7}
U_{o}(\theta,t)|0\rangle=d(t)\left(
\begin{array}{ccc}
-\cos\theta\bar{u}_{21}^{*}(t) \\
\bar{u}_{11}^{*}(t) \\
i\sin\theta\bar{u}_{21}^{*}(t) \\
\end{array}
\right).
\end{equation}
It is interesting to construct two reasonable functions for $\Theta(t)$:
\begin{equation}\label{eq8}
\Theta_1\left(t\right)= \kappa_1\sin^2(\frac{\pi}{T_1}t)
\end{equation}
and
\begin{equation}
\Theta_2\left(t\right)= \kappa_2\sin^2(\frac{\pi}{T_2}t) (1-\cos\lambda(t-T_2)),
\end{equation}
where $\lambda$ is defined as $D/2$, $\kappa_1$ and $\kappa_2$ are two parameters to be determined,  and $T_1$ and $T_2$ describe the time duration for the two cases.

Because the target state doesn't contain state $|0\rangle$, we need to set $\bar{u}_{11}(T_{1,2})=0$, {\it i.e.} $\cos[\frac{1}{2}W_s(\Theta_{1,2}, 2\lambda, T_{1,2})]=0$. Then the quantity $\chi(T_{1,2})=\lambda W_c(\Theta_{1,2},T_{1,2})$ contributes an overall phase in the state $U_o(\theta,T_{1,2})|0\rangle$ in Eq. (\ref{eq7}).
In order to achieve a minimal time value $T_{1,2}$ and a finite field pulse in the time interval $t\in(0, T_{1,2})$,
we need two conditions: $0<\chi(T_{1,2})\leq\frac{\pi}{2}$ and $W_s(\Theta_{1,2}, 2\lambda, T_{1,2})=\pi$.
Once we set $t=T_{1,2}$ and choose a value for $\chi(T_{1,2})$, the parameter $\kappa_{1,2}$
and time duration $T_{1,2}$ can be solved numerically. We can construct arbitrary superposed state\cite{Fang}
\begin{equation}
|q\rangle(\phi) =\cos\phi|+\rangle +i\sin\phi|-\rangle ~~~~ (0\leq\phi \leq \pi)
\end{equation}
in the following way. Choosing $\alpha$ and $\beta$ in Eq.(2) to satisfy the equality $\arctan(\beta/\alpha)=\theta \ge 0$, we can let $\theta = \pi-\phi$ in Eq. (7) and thus obtain the final state
\begin{equation}\label{eq10}
U_{o}(\pi-\phi,T_{1,2})|0\rangle=i\left(
\begin{array}{ccc}
\cos\phi\\
0  \\
i\sin\phi \\
\end{array}
\right)e^{-i\chi(T_{1,2})}e^{-i\frac{D}{2}T_{1,2}}.
\end{equation}
Neglecting the overall factor, we achieve the target state. If we set $\chi(T_{1,2})=\pi/2$, we can get a
self-consistent result: $\kappa_1=1.4627998$ and $T_1=2.42412/\lambda$, $\kappa_2=2.01024$ and $T_2=2.442705/\lambda$. The time-dependent function $\Theta_{1,2}(t)$ and the corresponding magnetic field $B_{1,2}(t)$ as functions of time $t$ in units of $T_{1,2}$ are shown in the Fig. 1. For the second function $\Theta_2(t)$, we use $1-\cos\lambda(t-T_2)$ to make the derivation of magnetic field at $T_2$ equal zero. The magnetic field is anti-symmetrical under the transformation $t\rightarrow 2T_2-t$ due to the symmetry of $\Theta_{1,2}(t)$. As we have pointed out\cite{Fang}, the arbitrary relative phase $\varphi$ can be realized by constant magnetic field and it has no contribution to the geometric phase, and hence we do not consider the relative phase.


{\it Nonadiabatic geometric Phases.} As usual, the geometric phase is defined as the difference between the total phase and the dynamical phase. We will consider three specific cases. In the first case, the initial state is $|0\rangle$. This is different from that for non-Abelian holonomic quantum computation\cite{Sjoq,CZu,Silvia}, because the dynamical phase is present during evolution and its geometric phase is independent of magnetic field direction. In the second case, when the initial state is superposed state $|q\rangle$, the phase evolution depends on magnetic field direction. We can still get a relatively simple result similar to the first one. In the third case, we allow the constant $\lambda$ to change and then get a very different result.

\begin{figure}
\begin{center}
\includegraphics[width=0.95\columnwidth]{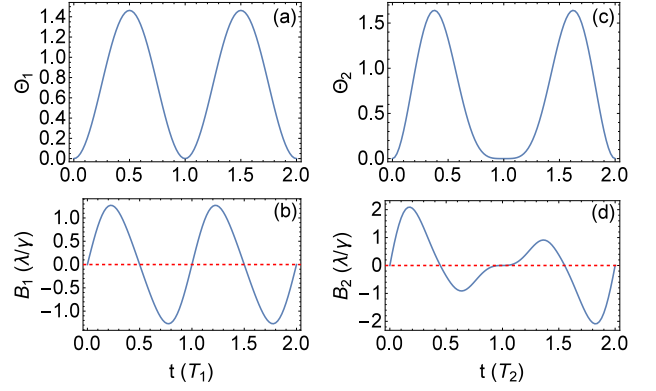}
\caption{\label{fig:qtest2} The time $t$ dependences of $\Theta_{1}(t)$ (a) and $\Theta_{2}(t)$, (c) and the corresponding  magnetic fields $B_{1}(t)$ (b) and $B_{2}(t)$ (d), with $\kappa_1$=1.4627998, $T_1\approx 1.689$ns, and $\kappa_2$=2.01024, $T_2\approx1.702$ns.}
\end{center}
\end{figure}


{\it Case 1.} The state begins from $|0\rangle$. The total phase can be calculated through $\langle 0|\psi(t)\rangle=e^{-i\lambda t}\bar{u}_{11}^{*}(t)$, and the dynamical phase by integrating $\Im\langle\psi_o(t)|\dot{\psi}_o(t)\rangle=\Im\langle0|U^{+}_{o}(\theta_1,t)\dot{U}_{o}
(\theta_1,t)|0\rangle$. We can express the geometric phase as
\begin{equation}\label{11}
\begin{aligned}
\varphi^{1}_{g}(t)=&\arg\langle 0|U_{o}(\theta_1,t)|0\rangle   \\
&-\Im\int_{0}^{t}\langle 0|U^{+}_{o}(\theta_1,t')\dot{U}_{o}(\theta_1,t')|0\rangle dt'.
\end{aligned}
\end{equation}
It can be proved that the integrand is equivalent to $-\lambda+\frac{\dot{\Theta}}{2}\sin(2\lambda W_c(\Theta,t))\sin W_s(\Theta, 2\lambda, t)+\lambda\cos\Theta\cos W_s(\Theta, 2\lambda, t)$. Here, we describe the magnetic field direction with $\theta_1$. Surprisingly, the geometric phase is independent of parameter $\theta_1$, and depends only on the parameter $\Theta(t)$.
The state $|0\rangle$ will evolve into its orthogonal state $|q\rangle(\pi-\theta_1)$ at time $T_{1,2}$. Considering that the total phase is not well-defined when the state is orthogonal, we can consider the phase at time $T_{1,2}$ as the left or right limit of $\varphi^{1}_{g}(t)$, i.e. $\varphi^{1}_{g}(T_{1,2})=\lim\limits_{t\rightarrow{T_{1,2}^{\pm}}}\varphi^{1}_{g}(t)$. When the system evolves to the time $t=2T_{1,2}$, the state will return to the initial state $|0\rangle$ with an overall phase just like the AA phase. Due to the symmetry of $\Theta_{1,2}(t)$, the probability evolution actually has the symmetry: $P_{|0\rangle}(t)=P_{|0\rangle}(2T_{1,2}-t)$. The phase evolution has a similar symmetry. Our numerical calculation shows that the sum of the phases at time $t$ and at $2T_{1,2}-t$ is constant. Surprisingly, we can show two equalities: $\arg\langle 0|U_{o}(\theta_1,t)|0\rangle + \arg\langle 0|U_{o}(\theta_1,2T_{1,2}-t)|0\rangle=\pi-2\lambda T_{1,2}$ and $\Im\int_{0}^{t}\langle\psi_o(t)|\dot{\psi}_o(t)\rangle+\Im\int_{0}^{2T_{1,2}-t}\langle\psi_o(t)|\dot{\psi}_o(t)\rangle=
\Im\int_{0}^{2T_{1,2}}\Im\langle\psi_o(t)|\dot{\psi}_o(t)\rangle$. As a result, the phase values between $T_{1,2}$ and $2T_{1,2}$ can be obtained from those in the region $[0,T_{1,2})$. With given $\Theta_{1,2}(t)$, the corresponding phase evolution and the probability evolution of state $|0\rangle$ are presented in Fig. 2. When the time changes from $t=0$ to $t=2T_1$ (or $2T_2$), the geometric phase is equivalent to $\pi$ (or $4.2705$) for $\Theta_{1}(t)$ (or $\Theta_{2}(t)$).

\begin{figure}
\begin{center}
\includegraphics[width=0.95\columnwidth]{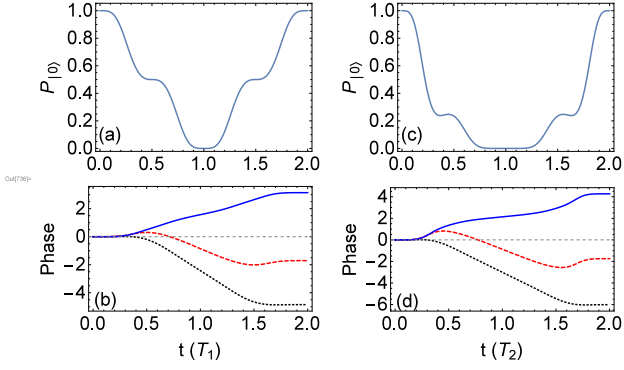}
\caption{\label{fig:qtest2} The time $t$ dependences of $P_{|0\rangle}$ for $\Theta_{1}(t)$ (a) and $\Theta_{2}(t)$ (c) starting from $|0\rangle$, and the corresponding evolution of the total phase (red dash), dynamical phase (black dot), and geometric phase (blue) (b, d). The geometric phases at $2T_{1}$ and $2T_{2}$ are $\pi$ and 4.2705, respectively.}
\end{center}
\end{figure}


{\it Case 2.} The initial state is a qubit state $|q\rangle$ normal to $|0\rangle$. It can be derived that $\langle q|U_{o}(\phi_1,t)|q\rangle=e^{-i 2\lambda t} \left(\sin ^2(\phi_1 +\phi)+\bar{u}_{11}(t) e^{i\lambda t} \cos ^2(\phi_1+ \phi)\right)$ and $\Im\langle\psi_q(t)|\dot{\psi}_q(t)\rangle=\Im\langle q|U^{+}_{o}(\phi_1,t)\dot{U}_{o}(\phi_1,t)|q\rangle$. Here, we mark the magnetic field direction as $\phi_1$. If $\phi_1+\phi=\pi$, the $|q\rangle(\phi)$ state will become $|0\rangle$ at $T_{1,2}$ and will return to the initial state $|q\rangle$ at the time $2T_{1,2}$.
The geometric phase can be expressed as
\begin{equation}\label{12}
\begin{aligned}
\varphi^{2}_{g}(t)=&\arg\langle q|U_{o}(\phi_1,t)|q\rangle   \\
&-\Im\int_{0}^{t}\langle q|U^{+}_{o}(\phi_1,t')\dot{U}_{o}(\phi_1,t')|q\rangle dt'.
\end{aligned}
\end{equation}
The integrand above can be proved to be equivalent to $-\lambda-\cos^2(\phi_1+\phi)(\lambda\cos\Theta\cos
W_s(\Theta, 2\lambda, t)+\frac{\dot{\Theta}}{2}\sin(2\lambda W_c(\Theta,t))\sin W_s(\Theta, 2\lambda, t))-\lambda \sin ^2(\phi_1+\phi)$. With the same magnetic field in Fig. 1, the probability evolution of state $|0\rangle$ and the geometric phase are presented in Fig. 3. Comparing Fig. 2 and Fig. 3, it can be seen that the geometric phase seems to be symmetric to each other in these two cases. It can be proved that they are actually the same except for a negative sign. The geometric phases gained at the ending times ($2T_1$ and $2T_2$) are equivalent to $-\pi$ and $-4.2705$ for $\Theta_{1}(t)$ and $\Theta_{2}(t)$, respectively.

\begin{figure}
\begin{center}
\includegraphics[width=0.95\columnwidth]{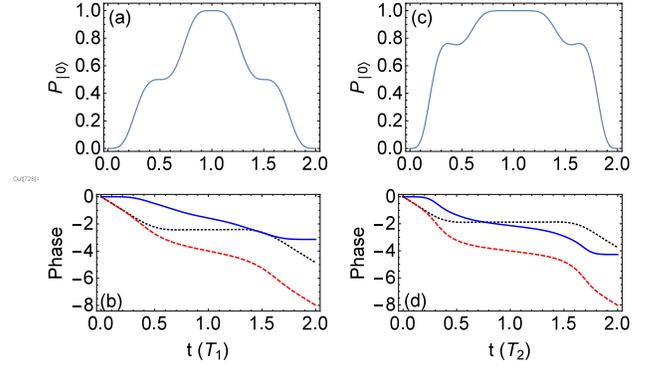}
\caption{\label{fig:qtest2} The time $t$ dependences of $P_{|0\rangle}$ for $\Theta_{1}(t)$ (a) and $\Theta_{2}(t)$ (c) starting from $|q\rangle$, and the corresponding total phase (red dash), dynamical phase (black dot), and geometric phase (blue) (b,d). The corresponding geometric phases at $2T_{1}$ and $2T_{2}$ are $-\pi$ and $-4.2705$, respectively.}
\end{center}
\end{figure}


{\it Case 3.} The parameter $\lambda$ in the function $\Theta_2(t)$ can be allowed to change. For convenience, we define $\eta = \lambda x$, where $x$ is a variable. Then we can construct the third function: $\Theta_3(t)=\kappa_3\sin^2(\frac{\pi}{T_2}t) (1-\cos\eta(t-T_2))$. Here, we set the time duration is the same as $\Theta_2(t)$, and $\kappa_3$ and $\eta$(or $x$) are adjustable parameter. The quantity $\chi(T_2)=\lambda W_c(\Theta_3, T_2)$. In order to construct a reasonable magnetic field easily, we must set the quantity $\chi(T_2)\leq\frac{\pi}{4}$ because of $\cot$ function in the magnetic field formula. We set it as $\frac{\pi}{4}$, and we need to solve the  self-consistent equation: $\chi(T_2)=\frac{\pi}{4}$ and $W_s(\Theta_3, 2\lambda, T_2)=\pi$. Consequently, we obtain a reasonable result: $\kappa_3=1.551569$ and $x=2.8671219$. At this time, there is no simple relation between time $t$ and $2T_2-t$ because of different phase conditions. In Fig. 4 we present the magnetic field and the probability evolution of state $|0\rangle$, and the phases (with the two initial states: $|0\rangle$ and $|q\rangle$) as functions of time $t$. The curves of the phase evolution with $x=1$ and $x=2.8671219$ are similar. The small oscillation in the curves is due to the additional function factor. The geometric phases gained between $t=0$ and the ending times ($2T_2$) are equivalent to $4.54$ and $-4.54$ with $|0\rangle$ and $|q\rangle$ as the starting states, respectively.

\begin{figure}
\begin{center}
\includegraphics[width=0.95\columnwidth]{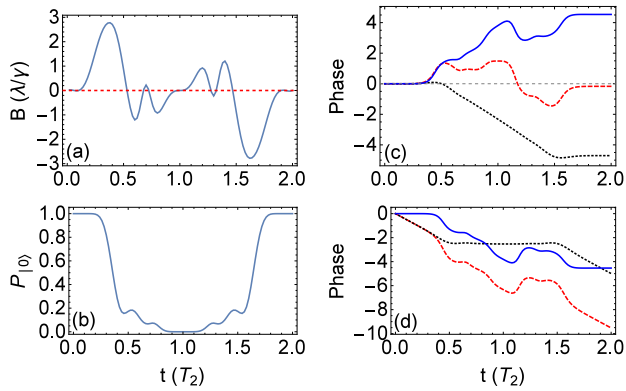}
\caption{\label{fig:qtest2} The time $t$ dependences of the magnetic field (a) and the probability $P_{|0\rangle}$ (b) starting from $|0\rangle$, and the corresponding total phases (red dash), dynamical phases (black dot), and geometric phases (blue) starting from $|0\rangle$ (c) and $|q\rangle$ (d). The geometric phases at $t=2T_{2}$ are $4.54$ (c) and $-4.54$ (d).}
\end{center}
\end{figure}


{\it Geometric interpretation. } We use the case of $\Theta_1(t)$  to understand the idea of geometric property. We rewrite the  equation(\ref{eq7}) as the following expression by introducing three real time-dependent functions: $\eta_s(t)$, $\theta_s(t)$, and $\varphi_s(t)$.
\begin{equation}
|\Psi(t)\rangle=e^{i\eta_s(t)}(\cos\frac{\theta_s(t)}{2}|0\rangle+e^{i\varphi_s(t)}\sin\frac{\theta_s(t)}{2}|q\rangle)
\end{equation}
In the time domain $(0,2T_1)$, the function $\theta_s(t)$ takes values in $(0, \pi)$, and the function  $\varphi_s(t)$ in the domain $(-\frac{\pi}{2},\frac{3\pi}{2})$. At $t=0$ the state is $|0\rangle$, and at $t=T_1$ the state becomes $|q\rangle$; and at $t=2T_1$ the state returns to $|0\rangle$. Therefore, we obtain a cyclic process and thus the path of the state in the unit sphere spanned by the ($\theta_s$,$\varphi_s$) parameters makes a closed curve. Then we can obtain the solid angle for the time duration $(0,2T_1)$: $|\int\int \sin \theta_s d\theta_sd\varphi_s|=2\pi$.
Our geometric phase $\pi$ in this case is equivalent to half the solid angle, as it should be. It is interesting that the geometric interpretation of the quantum phase is still true, although the quantum process is nonadiabatic.


{\it Conclusion.} It is believed that the study of geometric phases is an attempt to understand quantum mechanics better. Geometric phase is an observable quantity in the experiment with a solid-state spin qubit via spin echo interferometry\cite{Zhang,Leek,Yale}. It has been proposed to measure Berry phase in mechanically rotating diamond crystal\cite{Maclaurin}. Geometric phase has many potential applications\cite{Ledbetter,Ajoy}. With the development of quantum technique, geometric quantum computation makes a hot research field in quantum physics because of its fault tolerance property\cite{Liang,S,Tan}, and  it will be relevant to qubit control.
Our main results include: (1) we have constructed exact evolution operator of the NV spin system with the more transparent method, and then found three physically reasonable pulses for designing fast quantum logic gates based on the NV spin in diamond; and (2) we have investigated nonadiabatic geometric phases of the fast-driven NV spin systems, and shown that for the first pulse ($\Theta_1(t)$), the nonadiabatic geometric phase for the cyclic path ($t=0\rightarrow T_1\rightarrow 2T_1$) is equivalent to half the solid angle spanned by the corresponding two angle variable in the definition of the qubit state. In addition, the controlling pulse and the geometric phase can be manipulated through taking different $\Theta(t)$ function. We believe that these manipulations can be useful in designing practical quantum applications, and the nonadiabatic geometric phases, measurable in future experiments, could be used in quantum applications.

\begin{acknowledgments}
This work is supported by the Nature Science Foundation of China (Grant No. 11574366), by the Strategic Priority Research Program of the Chinese Academy of Sciences (Grant No.XDB07000000), and by the Department of Science and Technology of China (Grant No. 2016YFA0300701).
\end{acknowledgments}

\end{document}